# Nonequilibrium Magnetization of a Two-Dimensional Electron Gas in a Static Magnetic Field

D. R. Faulhaber and H. W. Jiang

Department of Physics and Astronomy, University of California at Los Angeles 405 Hilgard Avenue, Los Angeles, Ca 90095, USA

Using a sensitive DC torque magnetometer we measure the orbital magnetization by sweeping the density at fixed magnetic fields on GaAs-AlGaAs heterostructures. At low temperatures strong nonequilibrium magnetization signals dominate the data. In literature the observation of nonequilibrium signals are often associated with eddy currents generated by sweeping the magnetic field. The elimination of a changing magnetic field then poses a question regarding the origin of these signals. Our data suggests another source of nonequilibrium magnetization potentially due to the change of area occupied by compressible and incompressible states as one sweeps the Fermi energy from one Landau level to the next.

Magnetization measurements on a two dimensional electron system (2DES) subjected to strong perpendicular magnetic fields have revealed exciting new physics recently. Equilibrium magnetization measurements, manifested by the well known de Haas van Alphen effect (dHvA) have uncovered information on the broadening of Landau levels, the functional form of the density of states and gaps in the energy spectrum. This effect has been extensively studied in many 2DES's [1, 2, 3]. At lower temperatures and smaller values of the filling factor a nonequilibrium magnetization accompanies and dominates the dHvA signal [4,5,6,7]. It is generally accepted, the changing magnetic field generates an electromotive force which produces these currents and hence the magnetization signal. This signal is further amplified by the near zero resistance state occurring at low temperatures and integer filling factors. Much has been discovered about these nonequilibrium currents, including; their relation to quantum Hall breakdown, current paths within the 2DES and different scattering mechanisms. However, most experiments use non-contacted samples, where the density of carriers is held constant and the magnetic field is swept. In our experiment the magnetic field is fixed and the carrier density is swept, eliminating any effect arising from dB/dt. Surprisingly, our experiments reveal analogous nonequilibrium effects present at low temperatures as with experiments done with non-contacted samples. Here we report on the temperature and density sweep rate dependence of these nonequilibrium signals and provide a new model to explain their generation.

In this communication we report the study of these nonequilibrium effects from gated samples where the density is swept at fixed magnetic field. We characterize both temperature and sweep rate dependence of the nonequilibrium signals. We argue measured magnetization consists of two components; a current arising from a charging effect between the gate and the 2DES and an additional component associated with the changing magnetic flux penetrating the 2D system. With a static magnetic field we direct our attention to the changing area occupied by incompressible states as one sweeps the carrier density from one Landau level to the next.

The magnetometer employed is modeled after a design by Wiegers et al.[8], with modifications for simultaneous transport measurement and is shown in the inset of figure 1. The maximum torque sensitivity obtained is a few times $10^{-14}$ N-m at a magnetic field of 1T. The samples used are molecular-beam-eptaxy-grown, $GaAs/Al_{0.3}Ga_{0.7}As$ heterostructures.

Electrical contacts to the 2DEG are made by diffusing indium on all four corners of a square $4x4mm^2$ surface. An aluminum gate is evaporated over the remaining surface. With the gate we apply a voltage and tune the density of carriers of the 2DEG. The zero gate voltage mobility is $4.1x10^5 cm^2/Vs$ and the density $n_{2D} = 6.1x10^{11} cm^{-2}$, as found from Hall resistance and Shubnikov-de Haas oscillations. In this geometry we are able to take transport and magnetization measurements simultaneously during the same run insuring proper knowledge of the QH states.

Experimental magnetization data for a range of temperatures from 4K to 300mK at an applied magnetic field of 5 Tesla are shown in figure 1, including both forward (solid) and reverse (dashed) sweep directions. At 4K we can make out only the dHvA effect for the features at position $\nu=2,4$ and 6. As the temperature is decreased the signals quickly become dominated by the nonequilibrium magnetization component. The amplitude of these oscillations increases greatly with decreasing temperature and at 300mK corresponds to more than 100 times the dHvA oscillations. At lower temperatures the signals also evolve from symmetric spike-like features centered on the Landau level, to a more asymmetric sawtooth-like shape with the peaks, or minima shifted away from the Landau level centers. The direction of the oscillations flips under sweep direction reversal unlike the dHvA contributions.

The appearance of this nonequilibrium magnetization is very similar to early observations arising from currents being generated as one sweeps the magnetic field through the nearly dissipation less quantum-Hall states [5,6,7]. With dB/dt = 0, any magnetization contribution arising from a changing magnetic field is eliminated. We can however account for a current being injected into the system from a charging effect between the gate and the gas as we sweep the gate voltage. We analyzed the circuitry of the device by modeling it as a distributed circuit. Assuming the majority of the current flows near the sample boundary with several current paths within the bulk itself[4] we get an estimate of a few times $10^{-13}$ J/T, compared to the experimentally realized magnetization of a few times $10^{-12}$ J/T. One problem with accepting this injected current as being the sole source of the observed magnetization features is the explicit asymmetric nature of the signals in figure 1. We expect the component generated by a constant current to effectively mimic the inverse resistivity and be symmetric about filling factor, and to have a maximum value when the system is in the lowest resistance state, occurring precisely at integer filling factor[9]. We explored and ruled out other possible mechanisms for the generation of the observed signals including mechanical and electrical coupling from the magnetometer itself. For example, if the observed signals were a result of a torque interaction effect as proposed by shoenberg[13], then all dHvA oscillations would be affected. In figure 1 we clearly see no hysteretic behavior in the $\nu = 6$ oscillation nor do we see any in the higher temperature data of $\nu = 2,4$.

In principle it is possible to separate out the magnetization component generated by this charging effect by taking the average of both directional sweeps. To first order the average of the sweeps would cancel this component out while the difference of the sweeps would isolate it. This same method was adopted from reference 4. We separate two types of behavior in figure 2 using,

$$M_- = \frac{M_{reverse} - M_{forward}}{2}, \text{ and } M_+ = \frac{M_{reverse} + M_{forward}}{2}.$$

Figure 2A is our approximate charging component, it is symmetric about filling factor with a local maxima occurring at integer filling factor where the resistance of the state is a minimum. Since magnetization is proportional to the current which is inversely related to the resistance, we expect the maxima to be located here. We believe the increase in magnitude of the signal with decreasing temperature is a reflection on the changing resistance of the system as it moves closer to zero. Figure 2B on the other hand looks completely unfamiliar. We note a temperature independent point occurring at even integer filling factor, and analogous magnitude dependence with temperature as with Figure 2A. As of yet we can only explain the results from figure 2A as being associated with the charging effect.

In figure 3A we present the sweep rate dependence at a temperature of 1.5K, and clearly see the similarity with the temperature sweeps of figure 1. Similar data was taken at 300mK and we remark here, it is practically impossible to eliminate the nonequilibrium signal occurring at even integer filling factors and maintain experimental parameters, as the longest scan would take on the order of tens of hours to complete. In figures 3B, C we have again taken the average and difference of the data, presented as M₋ and M₊ for T= 1.5K and B = 5T. With our present results it is clear there is something in the data which is not accounted for by assuming merely the presence of charging currents. It is now our goal to describe how sweeping the gate voltage can generate the observed behavior in figure 3C.

In literature the nonequilibirum currents are driven by the changing magnetic field, having a static magnetic field then should eliminate these currents unless some sort changing area is occurring to our system as we sweep the density. To explain this qualitatively we model our system as consisting of a number of electrons in the 2D plane occupying a changing effective area which is induced by the density sweep. We assume currents travel in 1D channels throughout the bulk of the sample[4], with a magnetization contribution then given by:

$$M = IA = \frac{A}{R_{xx}} \frac{d\Phi}{dt} = \frac{AB}{R_{xx}} \frac{dA}{dt} = \frac{AB}{R_{xx}} \frac{dA}{d\upsilon} \frac{d\upsilon}{dt}$$

Here $I$ is the induced current, A is the area which is changing as a function of filling factor, $\Phi = BA$ is the magnetic flux and $d\Phi/dt = BdA/dt$ since $dB/dt = 0$, $R_{xx}$ is the resistance of the system, and $d\upsilon/dt$ is proportional to the experimental parameter $dV_g/dt$, the rate of sweep of the gate voltage. It is left then to discover how $AdA/dt$ changes with the applied gate voltage.

Our model is analogous to the percolation problem[10], where the potential $\varphi(x,y)$ on the 2D plane fluctuates due to the presence of disorder. These fluctuations divide the plane into filled and isolated regions. In other words, the areas occupied by electrons at any given Fermi energy are the incompressible regions, while the areas with electrons absent are the compressible regions. As the Fermi energy is swept towards integer filling the incompressible regions grow while the compressible regions shrink. When the Fermi energy finally reaches integer filling factor the entire structure is incompressible. Conversely, when the Fermi energy is precisely at a Landau level the entire structure is compressible. To model this changing compressible area, our first step is to find the profile of the potential arising from the donors on the 2D plane. To simplify things we use a simplified standard checkerboard potential of the form:

$$\Phi(x, y) = \Phi_o \sin\left(\frac{x+y}{2d}\right) \sin\left(\frac{x-y}{2d}\right). \quad [11]$$

Here, $\Phi_o$ is a constant which depends on the impurity parameters, and $d$ is their spacing.

Using this potential we can divide the 2D plane into regions of compressible and incompressible states. To do this we take a particular value of the gate voltage which is related to the Fermi energy and calculate how much area from the distribution is above or below this value. The area which lies above is the compressible (white) area while those below are the incompressible (black). In figure 4A we illustrate this point using an example of a random potential distribution. As the gate voltage is swept towards integer filling factor (hence the Fermi energy is swept) the incompressible regions grow and the compressible regions shrink. Electron states are located at the boundaries of these regions and electron orbit centers can be thought of as drifting on the equipotential lines circulating the perimeters [12]. In figure 4B we plot the compressible area A as a function of filling factor along with the best Gaussian fit to the data. The derivative $dA/d\nu$ is then taken numerically. The magnetization also contains a $1/R_{xx}$ dependence which we model as Gaussian peaks located on the LL centers. Since transport experiments are done simultaneously this approximation is compared directly and fits well. Putting this all together we arrive at a rough numerical calculation for $M = (AB/R_{xx})dA/dt$ plotted in figure 4C.

This simple calculation fits the experimental data in figure 3C surprisingly well. All of the features of the curve coincide with the experimental observations, including a sweep rate independent point occurring at integer filling, the sweep rate dependence, and the asymmetry of the peaks about the filling factor. In addition the magnitude of the numerical result is within 1.5 orders of magnitude of the experimental data, which we believe can be improved with a more rigorous calculation. Since the generated magnetization component depends inversely on $R_{xx}$, it is not surprising there is strong temperature dependence as illustrated in figure 2B. The resistance of the system changes greatly over the range of temperatures studied. A decrease in $R_{xx}$ then increases the magnitude of the magnetization signal, and vice versa. The magnitudes of both $M_+$ and $M_-$ have both been shown to depend on the $1/R_{xx}$ cofactor, which dominates the signals at lower temperatures. We believe it is this dependence which produces signals that are nearly equal in magnitude over the range of temperatures studied.

Most nonequilibirum experiments focus on the generation of a magnetization component which is driven by currents arising from $dB/dt \neq 0$. We have shown in the presence of a static magnetic field, a nonequilibrium magnetization signal can be generated while sweeping the 2D carrier density. This component depends directly on the area occupied by the compressible/incompressible states which changes as the density is swept from one

Landau level to the next. It is our belief this component should also be present in a dynamic field environment, superimposed with the signal arising from dB/dt = constant.

The authors would like to thank J. A. Robertson for helpful discussions and B. Alavi for technical assistance. This work is supported by NSF under grant DMR-0404445

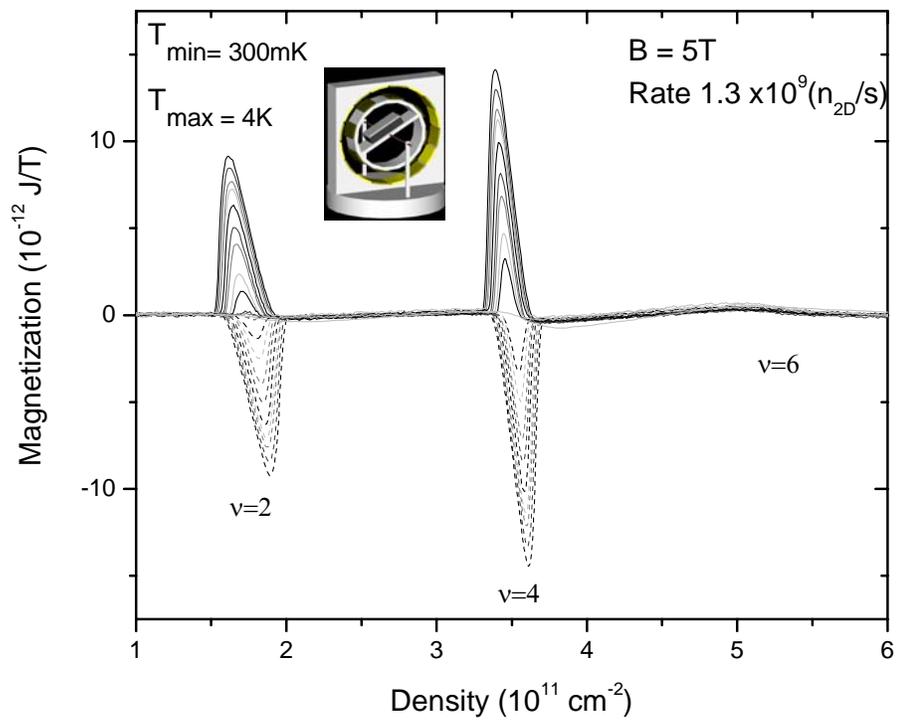

Figure 1) Experimental magnetization for temperatures 300mK, 400mK, 500mK, 600mK, 700mK, 800mK, 800mK, 900mK, 1.1K, 1.5K, 1.75K, 4Kat an applied magnetic field of 5T

and a sweep rate of 1.3 x 10$^9$ cm$^2$/s. Both forward (dashed) and reverse (solid) sweep directions are plotted. The positions of $\nu$ =2,4,6 are found from transport experiments taken simultaneously. Inset model of the torque magnetometer used for the experiment.

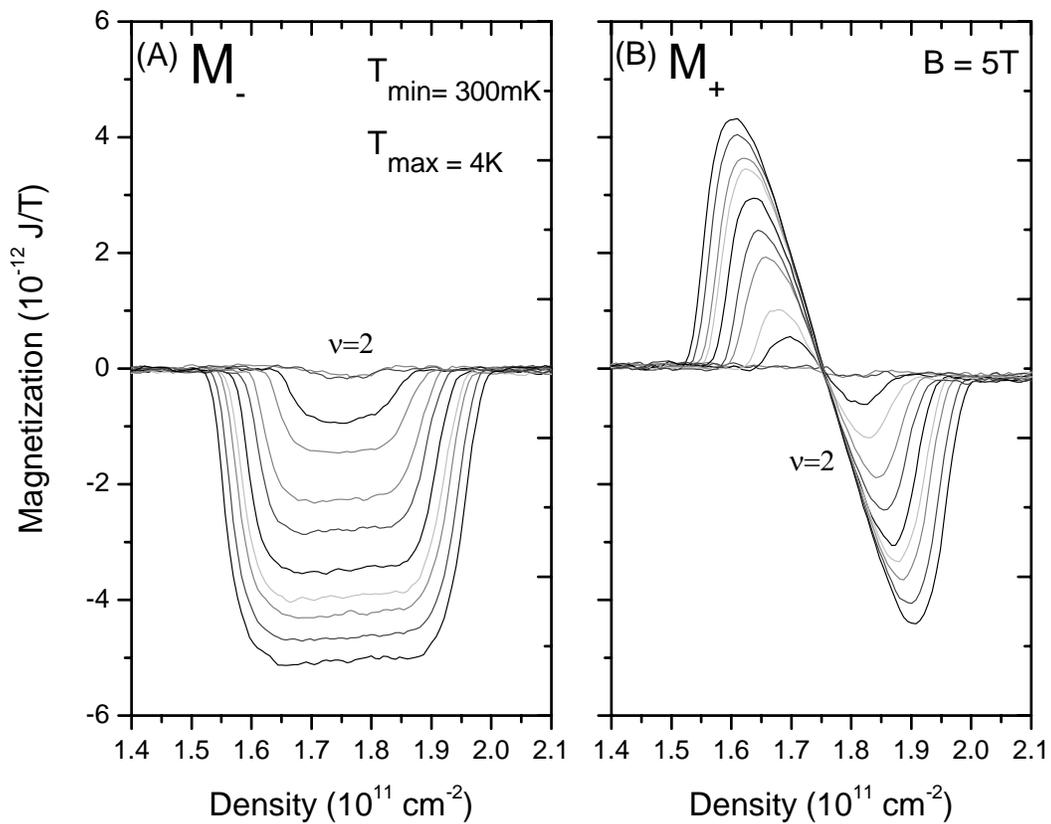

Figure 2) Here, we have separated the difference (M-) and average (M+) of sweep directions from figure 1. We plot only the feature at $\upsilon$ =2 for clarity, and note this position corresponds to a maximum magnetization amplitude in figure (A), and a temperature independent point in figure B. With the charging component being symmetric about the filling factor, to first order, M- represents the magnetization generated by a charging effect; while M+ eliminates this component.

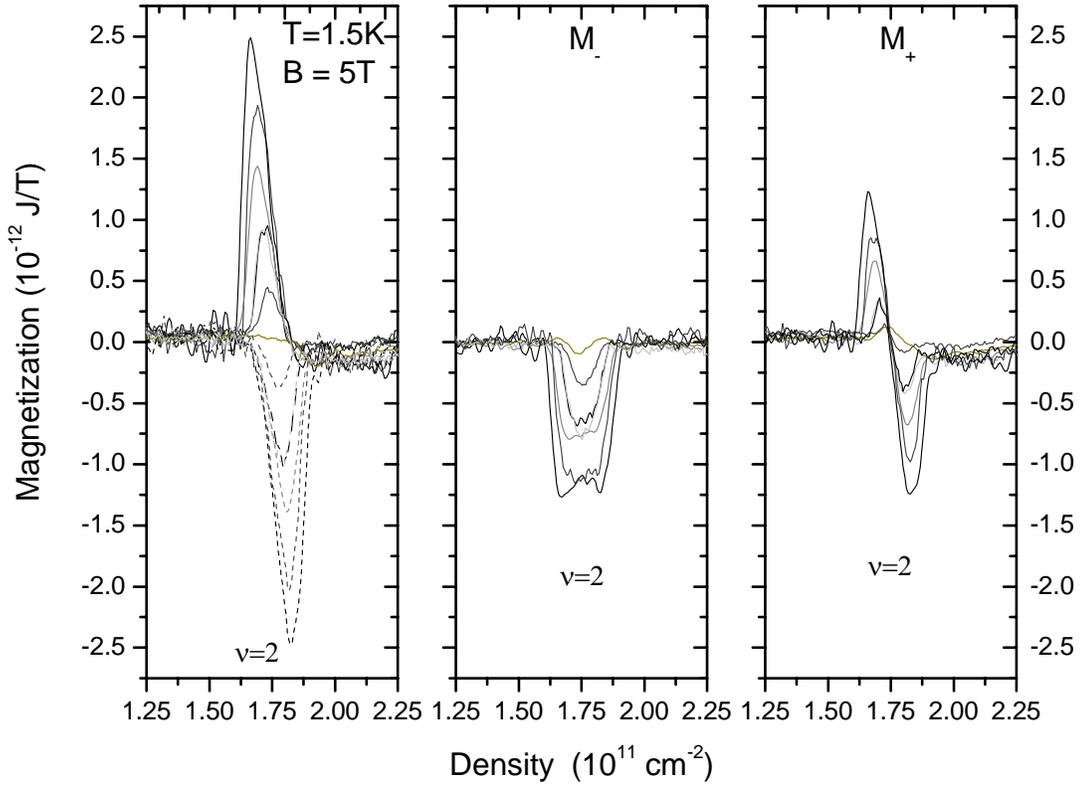

Figure 3) (A) Sweep rate dependence of magnetization signal occurring for $\nu = 2$ at an applied magnetic field of 5T, and T = 1.5K. Both forward (dashed) and back (solid) sweep directions are depicted. Gate sweep rates taken are .5V/s, .1V/s, .05V/s, .01V/s, .005V/s .0025V/s, .001V/s. The conversion to density sweep rate is .05V/s = $1.3 \times 10^9$ $n_{2D}cm^2$/s (B)

The difference of sweep directions which we attribute to a charging effect. (C) The average of the sweep directions.

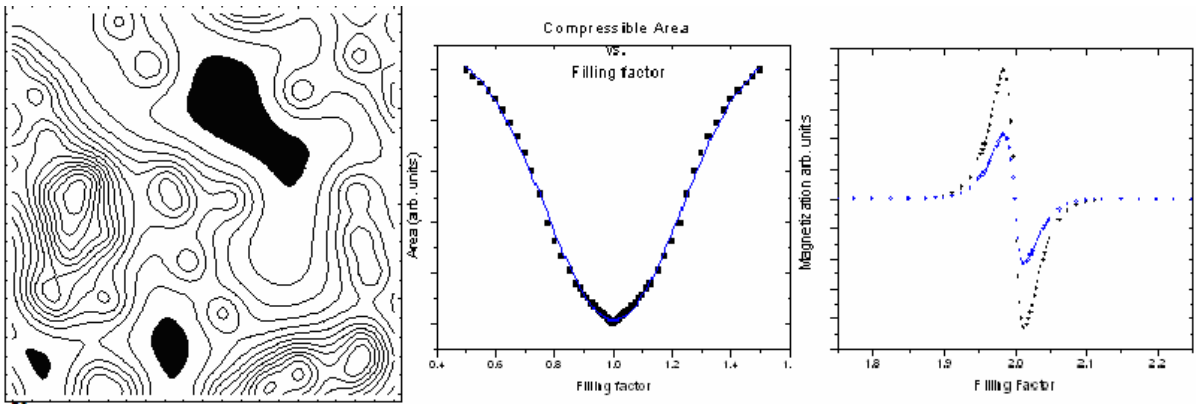

Figure 4) (A) An illustration of a more realistic potential profile with area occupied by compressible states (white) and incompressible states (black) when the Fermi energy lies close to a LL center. Contour lines represent equipotenial curves(B) The compressible area as a function of filling factor as calculated from our model using the checkerboard potential. The point at integer filling is discontinuous so we have taken the best Gaussian fit to the date to simplify further calculations. (C) The magnetization contribution from calculating this rate

of change of area as one sweeps the Fermi energy from one LL to the next, $M = -(AB/R_{xx})\, dA/dt$. Here we show two values of the sweep rate to show how the signal evolves.